\documentclass[reprint, 10pt, a4paper, aip, floatfix, longbibliography]{revtex4-2}
\usepackage[utf8]{inputenc}
\usepackage{graphicx} 
\usepackage{amsmath, amsbsy}
\usepackage{amssymb}
\usepackage{float} 
\usepackage{wrapfig} 
\usepackage[hidelinks]{hyperref}
\usepackage{tikz}
\usetikzlibrary{positioning}
\usetikzlibrary{shapes.misc,shadows}
\usetikzlibrary{decorations.markings}

\usepackage{enumerate}
\usepackage[shortlabels]{enumitem}

\begin{document}

\title{Wave-Supported Hybrid Fast-Thermal p-$^{11}$B Fusion}
\author{E.~J.~Kolmes}
\thanks{Co-first author}
\email{ekolmes@princeton.edu}
\author{I~E.~Ochs}
\thanks{Co-first author}
\email{iochs@princeton.edu}
\author{N.~J.~Fisch}
\affiliation{Department of Astrophysical Sciences, Princeton University, Princeton, New Jersey 08544, USA}
\date{\today}

\begin{abstract}
The possibility of fusion ignition in proton-Boron$^{11}$ plasma  is strongly enhanced if the energy from the fusion-produced $\alpha$ particles is channeled to fast protons, but in an environment in which most of the protons are thermally distributed.  
This  hybrid of thermonuclear fusion and beam-plasma fusion offers surprisingly large  advantages to either purely thermonuclear or purely beam-plasma fusion, neither of which can by themselves significantly exceed the large bremsstrahlung radiation emitted by the proton-Boron$^{11}$ plasma.  The hybrid scheme has the potential to reduce the confinement time  of the reactants that is required to achieve ignition by an order of magnitude.

\end{abstract}

\maketitle

Because Coulomb collision cross-sections are so much larger than fusion cross-sections, fast ions tend to slow down much more quickly than they fuse.
As a consequence, the leading approaches to controlled nuclear fusion tend to be thermonuclear, with reactants having roughly Maxwellian velocity distributions.  
However, these approaches suffer two main disadvantages.
First, many pairs of fuel ions necessarily have comparatively low relative velocities, and thus do not contribute significantly to fusion yields. 
Second, due to the collisions, charged byproducts of the fusion reaction tend to heat electrons, producing radiation and plasma pressure, but no direct fusion power.
This combination of low reactivity and high radiation is particularly detrimental to thermonuclear proton-Boron$^{11}$ (p-$^{11}$B) fusion, \cite{Rider1995, Rider1995a, Rider1997, Nevins2000CrossSection, Sikora2016CrossSection} which consequently has a narrow ignition window even under very optimistic assumptions, \cite{Putvinski2019, Cai2022TokamakPB11} requiring incredibly long energy confinement times---on the order of 500 seconds at ion densities $n_i\sim 10^{14}$ cm$^{-3}$.

The high electron radiation can, in principle, be ameliorated through the alpha channeling effect, where waves transfer power from fusion-born $\alpha$ particles directly into  fuel ions, 
thus bypassing  electrons and leading to enhanced fusion relative to radiation power. \cite{Fisch1992, Fisch1994, Fisch1995a, Herrmann1997, Hay2015ignition, Ochs2015a, Ochs2015b, Cianfrani2018, Cianfrani2019, Romanelli2020, White2021}
The premise of alpha channeling is that a wave can set up a diffusion path in phase space connecting higher energies at one spatial location to lower energies at another. 
If the population of particles on the higher-energy side exceeds that on the lower-energy side, then the wave-induced diffusion can result in a flow of fusion products from higher- to lower-energy regions of phase space. 
This transfers energy to the wave, which can then damp on ions. 
Since the p-$^{11}$B reaction, as opposed to the DT (deuterium-tritium) reaction,  is aneutronic, the $\alpha$ particles contain 100\% of the fusion power, so alpha channeling offers an even greater opportunity.

The  alpha channeling effect also addresses the low fusion reactivity of p-$^{11}$B if the  $\alpha$ particle energy is selectively channeled  to superthermal rather than thermal  protons.
Like in beam-target or beam-beam fusion approaches, \cite{Rostoker1997, Lampe1998, Labaune2013}  the fast-proton energy can be  chosen near the fusion reactivity maximum.
However, in beam-target fusion, the fusion power output never exceeds the power invested in supporting the beam against slowing down. \cite{Moreau77}
The beam-target approach can be improved if fusion products selectively energize high-energy fuel ions through knock-on collisions, \cite{Eliezer2016,Shmatov2016,Eliezer2020,Belloni2021} in a process with similar results to the alpha channeling effect; however, since there is no resonance condition, there is less control over which fuel ions are energized. 

Surprisingly, however, it turns out that in an alpha channeling-supported plasma, neither purely beam fusion nor purely thermonuclear fusion is optimal.
Instead, a hybrid scheme is optimal, involving both an alpha channeling-supported minority fast proton beam and a majority thermal proton population.
The fusion power from the fast ions alone cannot support them against slowing down on the thermal population, so some thermal population is required to support the fast ions. 
At the same time, energy lost from the fast ions to the background thermal protons now stands to produce at least some fusion.
In fact, such an alpha-channeling-driven hybrid fusion scheme significantly reduces the fusion ignition requirement, achieving ignition while allowing five times as much power to be lost through thermal conduction relative to the thermonuclear case.

To demonstrate the advantage provided by a wave-supported hybrid approach, consider a  0D power balance model, neglecting synchrotron radiation.
Synchrotron radiation can be important, but it is highly configuration-dependent, can be reabsorbed within the plasma after emission, and can be reflected back into the plasma.\cite{Bekefi, Nevins1998, Volosov2006ACT} Bremsstrahlung, on the other hand, is an essentially irreducible loss channel.

The  reactor is then characterized by three powers:
the fusion power, $P_F$, which heats the plasma; and the bremsstrahlung radiation $P_B$ and thermal conduction loss $P_L$, which cool it.
Balancing inputs and outputs:
\begin{align}
	P_F - P_B - P_L = 0. \label{eq:PLConstr}
\end{align}
Both $P_F$ and $P_B$ are determined by  the local plasma parameters, while $P_L$ is determined by the device specifics.
However, $P_L$ is related to the energy confinement time by $P_L = E / \tau_E$, 
where $E$ is the confined energy of the plasma constituents and $\tau_E$ is the energy confinement time.  
Though different figures of merit can be used, \cite{WurzelHsu2022} we choose to minimize the Lawson criterion $n_i \tau_E$, 
where $n_i$ is the ion density. \cite{Lawson1957}
For a given ion density,
we  then   maximize $P_L = P_F - P_B$,
thereby to minimize $\tau_E$, and so minimize the required energy confinement efficiency.
Note that the results reported here do not require a device that can actually attain the resulting $\tau_E$.
A device with lower confinement times could behave in exactly the way we describe here, if it were held in steady state by auxiliary heating (depending on which species was being heated). 

Fast protons $f$ and thermal protons $p$  produce different amounts of fusion power. 
Consider, for simplicity,  fixed energy for the fast protons,  fixed boron density $n_b$, and fixed total proton density $(n_f + n_p)$.
Let $\phi \doteq n_f / (n_f + n_p)$. 
If $y_f$ and $y_p$ are the fusion power densities that would result if all protons were fast or thermal, respectively, then
\begin{align}
P_F = \phi y_f + (1-\phi) y_p . \label{eqn:PF}
\end{align}
To maximize fusion power minus bremsstrahlung radiation, note that the radiation depends on the electron temperature, which depends on the power balance for each species.
In terms of the total energy $U_s$ contained in each species $s$, 
\begin{align}
&\frac{dU_f}{dt} = \alpha_f P_F - (\kappa_p + \kappa_b + \kappa_e) \phi \\
&\frac{dU_p}{dt} = \alpha_p P_F + \kappa_p \phi + K_{pb} (T_b - T_p) \nonumber \\
&\hspace{30 pt}+ K_{pe} (T_e - T_p) - \gamma_p P_L \\
&\frac{dU_b}{dt} = \alpha_b P_F + \kappa_b \phi + K_{pb} (T_p - T_b) \nonumber \\
&\hspace{30 pt}+ K_{be} (T_e - T_b) - \gamma_b P_L \\
&\frac{dU_e}{dt} = \alpha_e P_F + \kappa_e \phi + K_{pe} (T_p - T_e) \nonumber \\
&\hspace{30 pt}+ K_{be} (T_b - T_e) - \gamma_e P_L - P_B. 
\end{align}
The $\alpha_s$ parameters describe the transfer of  $\alpha$-particle power to each particle species (either by alpha channeling \cite{Fisch1992} or by collisional slowing-down). 
The $\kappa_s$ parameters describe the rate at which the fast protons slow down on  other species, with $\kappa \doteq \kappa_p + \kappa_b + \kappa_e$; the $K_{ss'}$ parameters describe temperature equilibration between species; and the $\gamma_s$ parameters describe the fraction of the non-bremsstrahlung losses  sustained by each species. 

Since the ion-ion temperature equilibration rate tends to be larger than the other characteristic rates, taking $T_p = T_b$
substantially simplifies these equations. 
A further useful simplification is taking $\gamma_e = 0$, representing the (pessimistic) limit in which the confinement losses are all sustained by the ions rather than the electrons. 
Nonzero $\gamma_e$ would make the power balance more favorable by cooling the electrons. 
These simplifications lead to the following system of equations: 
\begin{align}
&\frac{dU_f}{dt} = \alpha_f P_F - \kappa \phi \label{eq:dUfdt}\\
&\frac{dU_i}{dt} = \alpha_i P_F + \kappa_i \phi + K_{ie} (T_e - T_i) -  P_L  \label{eq:dUidt}\\
&\frac{dU_e}{dt} = \alpha_e P_F + \kappa_e \phi + K_{ie} (T_i - T_e)  - P_B. \label{eq:dUedt}
\end{align}
Here $K_{ie} \doteq K_{pe} + K_{be}$ and $\kappa_i \doteq \kappa_p + \kappa_b$. 
In steady state, summing these equations yields Eq.~(\ref{eq:PLConstr}), so that Eq.~(\ref{eq:PLConstr}) can replace Eq.~(\ref{eq:dUidt}).
Importantly, in contrast to earlier works which examined alpha channeling in p-$^{11}$B plasmas, \cite{Hay2015ignition} we retain thermal conduction losses.

\begin{figure*}
    \centering
    \includegraphics[width=.7\linewidth]{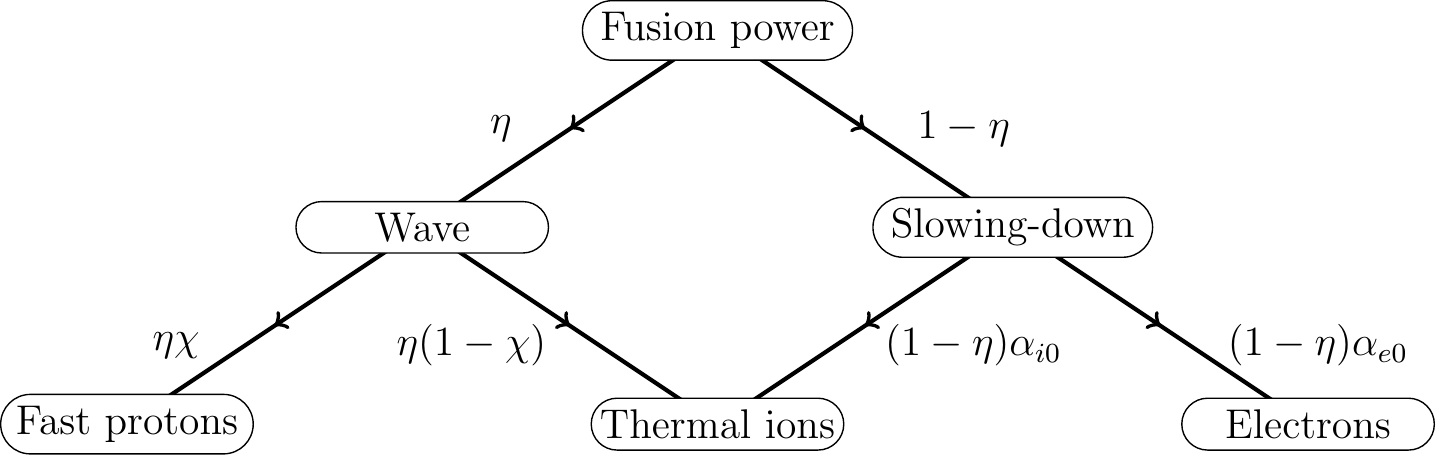}
    \caption{This figure shows how the fusion power is distributed between the different particle species, both through wave-facilitated channels (alpha-channeling) and through collisional slowing down. }
    \label{fig:powerFlow}
\end{figure*}

Alpha channeling can be modeled through modification of the $\alpha_s$ parameters. 
To measure the utility of alpha channeling, 
let $\eta$ be the overall alpha channeling efficiency, i.e., the fraction of $\alpha$-particle power that is transferred from $\alpha$-particles into the wave.
Then, to measure the relative utility of fast vs.~thermal protons, define a parameter $\chi$, representing the fraction of the channeled power directed into the fast protons, with the remaining fraction $(1-\chi)$ going into the thermal ions (Fig.~\ref{fig:powerFlow}).
Then
\begin{align}
\alpha_f &= \eta \chi \\
\alpha_i &= \eta (1-\chi) + (1-\eta) \alpha_{i0} \\
\alpha_e &= (1-\eta) \alpha_{e0},
\end{align}
where $\alpha_{s0}$ is the fraction of the $\alpha$-particle power that would be absorbed by species $s$ in the absence of alpha channeling (by collisional slowing-down). 
In the absence of alpha channeling, fast particles will tend to primarily slow down on the boron ions, so it is a reasonable approximation to ignore the $\phi$-dependence of $\alpha_{i0}$ and $\kappa_i$ (the latter of which would enter through the dependence of $\kappa_p$ on the thermal proton density). 

Thus, left with 6 variables ($\eta$, $\chi$, $\phi$, $P_L$, $T_i$, and $T_e$) and 3 constraints (Eqs.~(\ref{eq:PLConstr}), (\ref{eq:dUfdt}), and (\ref{eq:dUedt})), 
we are free to choose 3 variables, with the others determined by the constraints.
To show the fundamental advantages inherent in alpha channeling, while retaining analytic simplicity, we leave $T_i$ fixed, optimizing $P_L$ over $\eta$ and $\chi$. 

With $\eta$ and $\chi$ the independent variables, and $T_i$  fixed, we find $\phi$  from Eqs.~(\ref{eq:dUfdt}) and (\ref{eqn:PF}): 
\begin{align}
	0 &= \eta \chi [\phi y_f + (1-\phi) y_p] - \kappa \phi,
\end{align}
giving
\begin{align}
	\phi = \frac{\eta \chi y_p}{\kappa - \eta \chi (y_f - y_p)} = 1 - \frac{\kappa - \eta \chi y_f}{\kappa-\eta \chi (y_f - y_p)} \, . \label{eqn:phiSolution}
\end{align}

\begin{figure}[b]
    \centering
    \includegraphics[width=\linewidth]{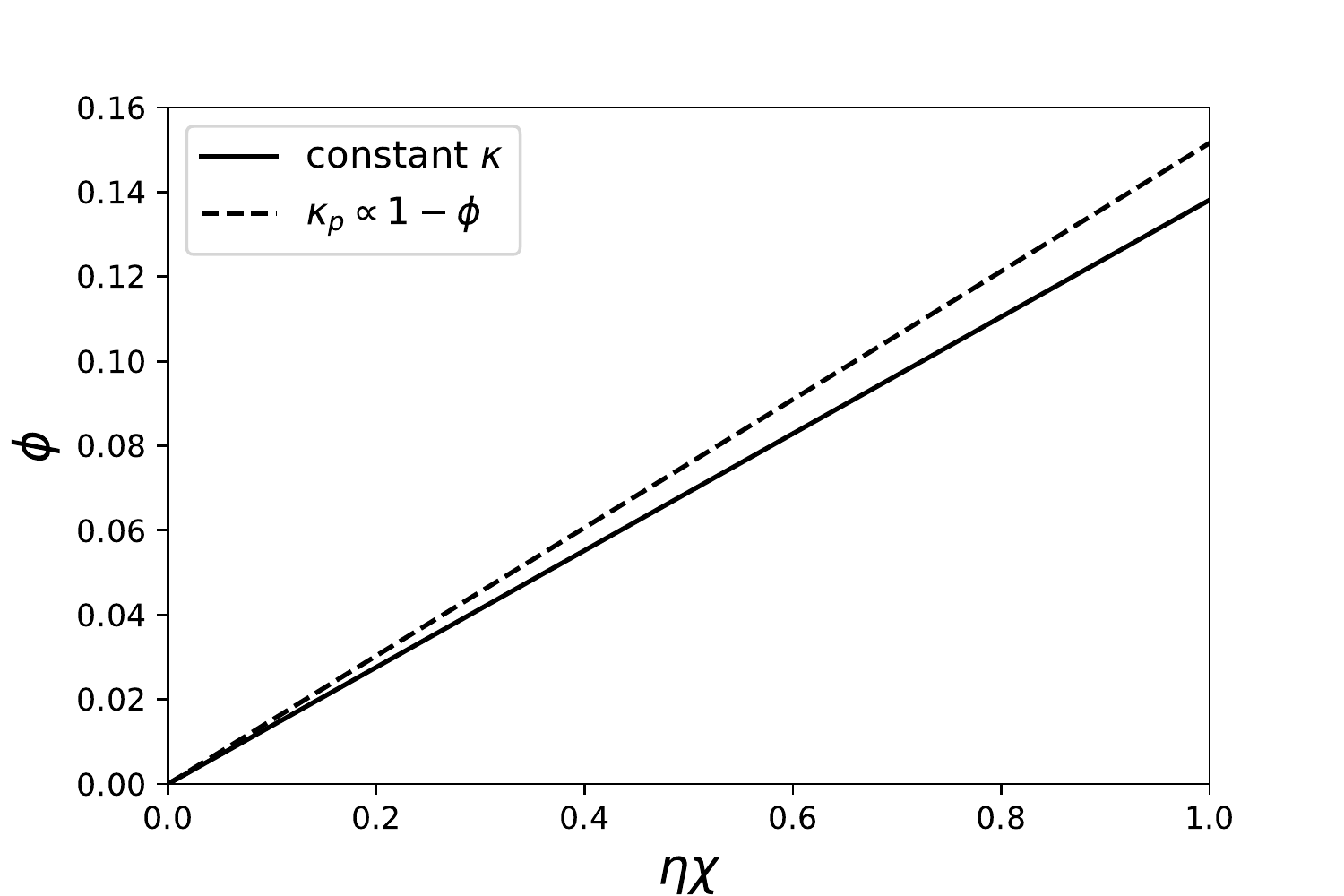}
    \caption{This figure shows $\phi$ as a function of $\eta \chi$ for a scenario in which all thermal species have a temperature of 300 keV, the fast protons have an energy of 643 keV, the total proton density is $8.5 \times 10^{13} \text{ cm}^{-3}$, and the boron density is $1.5 \times 10^{13} \text{ cm}^{-3}$. The dashed line shows the result when the $\phi$-dependence of $\kappa_p$ is retained. }
    \label{fig:phi}
\end{figure}

This solution is plotted for a characteristic set of parameters in Fig.~\ref{fig:phi}. 
With Eq.~(\ref{eqn:phiSolution}) in hand, the optimal beam fraction $\phi$ follows directly from the optimal choices of the channeling parameters $\eta$ and $\chi$. 
However, we can already derive analytically a fundamental result.
Recall that $\eta$ is an efficiency, and $\chi$ is a fraction of the alpha particle power, so
\begin{align}
    0<\eta<1; \quad 0< \chi < 1. 
\end{align}
Additionally, we know that the beam cannot support itself against slowing down, even at 100\% alpha channeling efficiency, which means that $\kappa > y_f$.
Finally, we know that the beam has a larger fusion cross section than the thermal population, so $y_f > y_p > 0$.
From the second half of Eq.~(\ref{eqn:phiSolution}), we see that, taken together, these parameter constraints mean that:
\begin{align}
	0 \leq \phi < 1,
\end{align}
with equality holding only if $\eta \chi = 0$.
Thus, if it turns out to be optimal to channel energy into the fast ions -- that is, if $P_L$ is maximized for $\chi \neq 0$ -- this implies that \emph{the optimal mix of protons is neither fully thermal nor fully fast protons, but a mix of the two.}
In other words, it follows immediately that the optimization must reside in a hybrid  of both beam-target and thermal schemes.

Now to proceed with the optimization, consider the choice of $\eta$, at a fixed value of $\chi$. 
Viewing $T_e$ and $\phi$ as functions of $\eta$ and $\chi$ determined by the constraint equations (\ref{eq:dUedt}) and (\ref{eqn:phiSolution}), we have:
\begin{align}
\frac{\partial P_L}{\partial \eta} &= (y_f - y_p) \frac{\partial \phi}{\partial \eta} - \frac{\partial P_B}{\partial T_e} \frac{\partial T_e}{\partial \eta} \, . 
\end{align}
Note that $P_B$ is an increasing function of $T_e$. 
The derivative of $T_e$ can be obtained from Eq.~(\ref{eq:dUedt}). To a good approximation, $\alpha_e$, $\kappa_e$, and $K_{ie}$ all scale like $T_e^{-3/2}$. Thus: 
\begin{align}
\frac{\partial T_e}{\partial \eta} &\approx \frac{-\alpha_{e0} P_F + [\alpha_e (y_f - y_p) + \kappa_e] \partial \phi / \partial \eta}{K_{ie} + 3 P_B / 2 T_e + \partial P_B / \partial T_e} \, . 
\end{align}
This immediately tells us that, in the absence of fast ions, the best results are obtained by maximizing the alpha channeling: 
that is, when $\chi = 0$, $P_L$ is maximized for $\eta = 1$. 
To see this, note that when $\chi = 0$, $\phi$ and $\partial \phi / \partial \eta$ both vanish. 
Therefore, when $\chi = 0$, the sign of $\partial P_L / \partial \eta$ is always opposite that of $\partial T_e / \partial \eta$, and (again, when $\chi = 0$) $\partial T_e / \partial \eta$ is always negative. 
In the absence of a beam, increasing the alpha channeling efficiency $\eta$ at constant $T_i$ lowers the electron temperature $T_e$ and decreases the associated radiative losses. 

To understand the more general case when $\chi \neq 0$, it is necessary to calculate the derivative of $\phi$: 
\begin{align}
\frac{\partial \phi}{\partial \eta} &= \frac{\kappa \chi y_p}{[\kappa - \eta \chi (y_f - y_p)]^2} \bigg[ 1 + \frac{3 \eta}{2} \frac{\kappa_e}{\kappa} \frac{1}{T_e} \frac{\partial T_e}{\partial \eta} \bigg] . 
\end{align}
In principle, it could be possible to get $\partial \phi / \partial \eta < 0$. 
However, note that in the parameter regimes relevant for p-$^{11}$B fusion, fast ions slow down much more strongly on ions than on electrons (see, e.g., Ref.~\cite{Putvinski2019}). 
As a result, $\kappa_e / \kappa$ and $\alpha_e$ are both small.
Moreover, note that $y_f > y_p$, and that $\kappa > y_f$ (otherwise ignition would be possible without any thermal proton population). 
So long as $\kappa_e / \kappa$ and $\alpha_e$ are sufficiently small, and so long as $y_f - y_p$ is not too close to $\kappa$, it follows that both $\phi$ and $P_L$ are increasing functions of $\eta$. 
Intuitively, this is because a higher $\eta$ means diverting power away from directly heating the electrons, which tend to radiate away their energy, and because higher $\eta$ allows for a larger beam fraction and a correspondingly larger fusion power density.

It thus follows that, regardless of $\chi$, increasing the channeling efficiency $\eta$ results in larger $P_L$.
However, it remains to determine the optimal value of $\chi$, the  fraction of power going  into fast rather than thermal protons.

To optimize over $\chi$, first combine Eqs.~(\ref{eqn:PF}) and (\ref{eqn:phiSolution}):
\begin{align}
    P_F 
    &= \frac{\kappa y_p}{\kappa - \eta \chi (y_f-y_p)} \, .
\end{align}
To maximize, take the derivative with respect to $\chi$:
\begin{align}
	\frac{\partial}{\partial \chi} \big( P_F - P_B \big) &= \frac{\eta \kappa y_p (y_f - y_p)}{[\kappa - \eta \chi (y_f - y_p)]^2}  - \frac{\partial P_B}{\partial \chi} \, . \label{eqn:PFDerivative}
\end{align}
Note that if $T_i$ and the fast ion energy are fixed, then $y_f$ and $y_p$ are also fixed. 
It follows from Eq.~(\ref{eqn:phiSolution}) that 
\begin{gather}
\frac{\partial \phi}{\partial \chi} = \frac{\eta \kappa y_p}{[\kappa - \eta \chi (y_f - y_p)]^2} \, ,
\end{gather}
so the condition on the sign of the derivative is 
\begin{gather}
\text{sgn} \bigg[ \frac{\partial}{\partial \chi} \big( P_F - P_B \big) \bigg] = \text{sgn} \bigg[ y_f - y_p - \frac{\partial P_B}{\partial \phi} \bigg|_\eta \, \bigg]. 
\end{gather}
The $\phi$ derivative of $P_B$ is taken at constant $\eta$. 
If $P_B$ does not depend on $\phi$, then the maximum occurs at the edge of the domain with $\chi = 1$, and it is always better to channel more energy into fast protons.
Thus, even though the solution is characterized by a mix of fast and thermal protons, this does not represent an interior point of the optimization itself (which channels maximum power to fast protons), but is rather a consequence of the constraint of having to support the fast proton distribution against slowing down. 

More generally, $P_B$ will increase with $T_e$, and $T_e$ will depend on $\phi$. 
In principle, if $P_B$ grows sufficiently quickly when the beam fraction is increased, a higher $\chi$ (and a correspondingly higher $\phi$) might not always be more favorable. 
In order to estimate how $P_B$ will vary, compute the response of $T_e$ when the system's other parameters vary at fixed $\eta$. 
This can be accomplished (as before) by taking the $\chi$ derivative of Eq.~(\ref{eq:dUedt}): 
\begin{align}
&\bigg[ - \frac{\partial \alpha_e}{\partial T_e} - \frac{\partial \kappa_e}{\partial T_e} \phi - \frac{\partial K_{ie}}{\partial T_e} (T_i - T_e) + K_{ie} + \frac{\partial P_B}{\partial T_e} \bigg] \frac{\partial T_e}{\partial \chi} \nonumber \\
&\hspace{80 pt}=\alpha_e \frac{\partial P_F}{\partial \chi} + \kappa_e \frac{\partial \phi}{\partial \chi} \, . \label{eqn:chiDerivative}
\end{align}
Again, $\alpha_e$, $\kappa_e$, and $K_{ie}$ are taken to scale like $T_e^{-3/2}$ and we neglect the $\phi$-dependence that appears in the transfer coefficients through the dependence on $n_p$. 
Then Eq.~(\ref{eqn:chiDerivative}) can be rewritten as 
\begin{align}
\frac{\partial T_e}{\partial \chi} &= \bigg( \frac{\partial \phi}{\partial \chi} \bigg) \frac{\alpha_e (y_f - y_p) + \kappa_e}{K_{ie} + \partial P_B / \partial T_e + 3 P_B /2 T_e} \, . 
\end{align}
Then, since $\partial P_B / \partial T_e > 0$, 
\begin{align}
&\text{sgn} \bigg[ \frac{\partial}{\partial \chi} \big( P_F - P_B \big) \bigg] = \nonumber \\
&\text{sgn} \bigg\{ (y_f - y_p) - \kappa_e \frac{\partial P_B}{\partial T_e} \bigg[ K_{ie} + \frac{3 P_B}{2 T_e} + (1-\alpha_e) \frac{\partial P_B}{\partial T_e} \bigg]^{-1} \bigg\} . \label{eqn:bremsCondition}
\end{align}
Here, the positive first term on the RHS corresponds to enhancement in fusion power due to the fast ions, while the negative second term corresponds to increase in the bremsstrahlung radiation due to fast ions heating the electrons. 
The expression is positive in the regimes of greatest interest for p-$^{11}$B devices. 
Consider, for example, a system with $T_i = 300 \text{ keV}$, $n_p = 8.5 \times 10^{13} \text{ cm}^{-3}$, and $n_b = 1.5 \times 10^{13} \text{ cm}^{-3}$, around the thermonuclear optimum described in Ref.~\cite{Putvinski2019}. 
For these parameters, $y_f - y_p$ is very close to $\kappa_e$ -- they are within 10\% of one another. 
The second (negative) term in Eq.~(\ref{eqn:bremsCondition}) is then substantially suppressed by the rest of the expression; $K_{ie}$, $P_B / T_e$, and $\partial P_B / \partial T_e$ are all roughly the same size. 
As a result, higher $\chi$ results in higher equilibrium $P_L$, so the optimal fast particle fraction $\phi$ is the highest one that the slowing-down power will permit. 
This is shown numerically in Fig.~\ref{fig:PLs}; the associated numerical calculation is discussed in Appendix~\ref{appendix:model}.

\begin{figure}
    \centering
    \includegraphics[width=\linewidth]{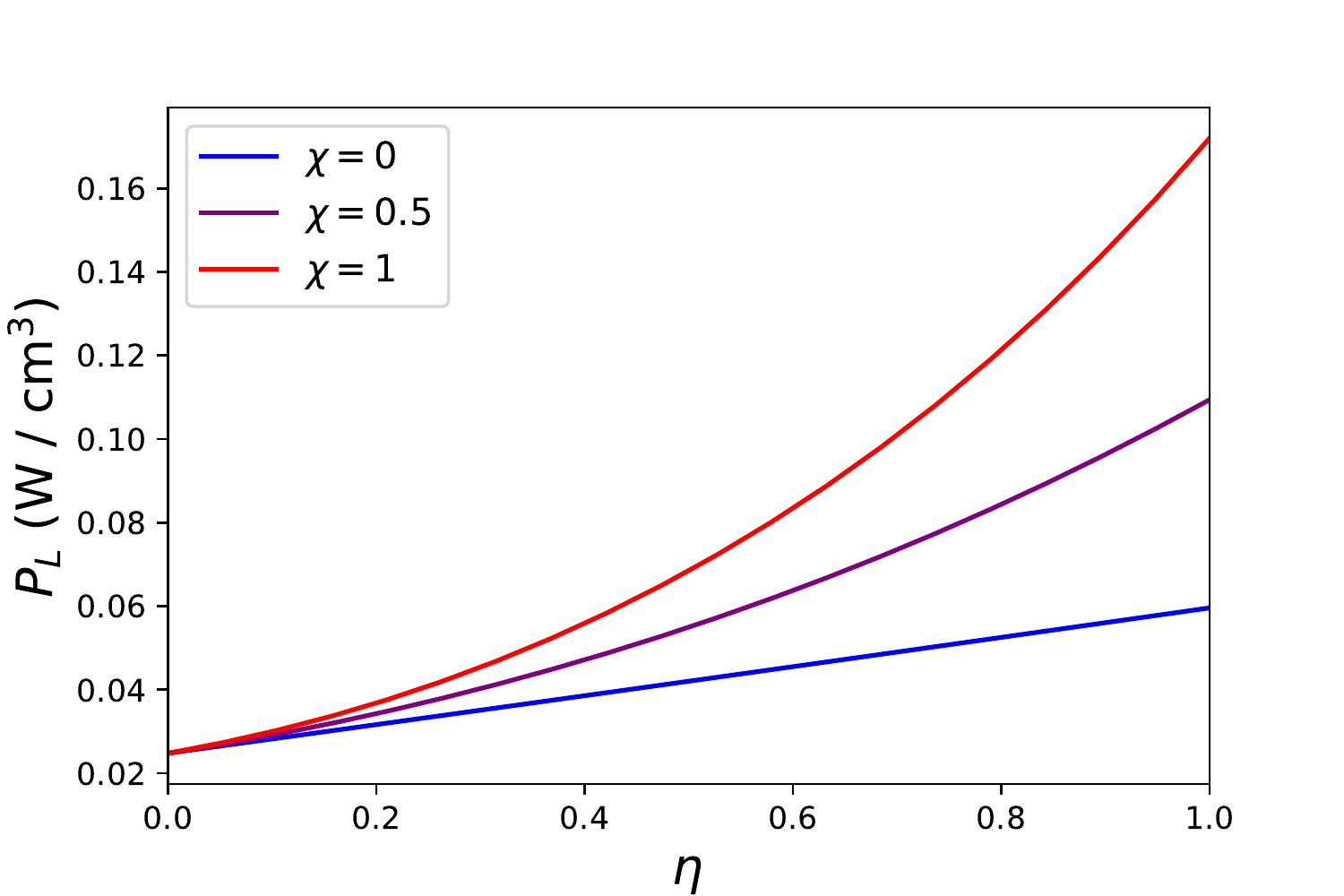}
    \caption{This figure shows $P_L = P_F - P_B$ as a function of $\eta$ for several choices of $\chi$, using $T_i = 300 \text{ keV}$, a fast proton energy of 643~keV, $n_p + n_f = 8.5 \times 10^{13} \text{ cm}^{-3}$, and $n_b = 1.5 \times 10^{13} \text{ cm}^{-3}$. Particularly for higher $\eta$, the larger choices of $\chi$ (which correspond to a larger fast particle fraction $\phi$) lead to higher $P_L$. In other words, in this parameter regime, it is favorable to channel as much power as possible into the fast proton population. The most favorable operating point involves a mix of fast and thermal protons, since $\chi = 1$ does not lead to $\phi = 1$.}
    \label{fig:PLs}
\end{figure}

With a 0D model of energy exchange and modest further approximations, we capture rigorously and succinctly the upside potential of employing alpha channeling in p-$^{11}$B plasma.
Assuming that the fusion power can be channeled to either thermal protons or fast protons near the fusion power maximum, we derived three rules for p-$^{11}$B fusion:
\begin{itemize}[leftmargin=0pt,label={}]
\item{\textbf{Rule 1:}}  The optimum arrangement lies in a hybrid approach of both thermal and fast protons.
\item{\textbf{Rule 2:}} The optimum arrangement lies in maximizing  the alpha channeling effect.
\item{\textbf{Rule 3:}} This  optimum  lies in channeling the $\alpha$-particle power solely to the fast ions, notwithstanding Rule 1.
\end{itemize}
These rules are not only valuable for navigating through the very large parameter space of possibilities for the plasma operating conditions, but they also point to a particularly promising opportunity: namely that, through alpha channeling, the allowable conduction loss $P_L$ for ignition might be increased by a factor of $\sim 3$ without a beam and by a factor of $\sim 5$ with a beam. 
Thus, we arrive at the remarkable result that the prospects for p-$^{11}$B fusion might be seriously improved in a {\it hybrid} approach, with alpha channeling transferring energy to fast protons, but in a predominantly thermal proton population.  
Without  alpha channeling, we recover the result of Putvinski, \cite{Putvinski2019} which reaches marginal conditions for ignition requiring extremely long confinement times.

The analysis  here captures the relevant physics as simply as possible.
It reveals the key features and general scalings that help alpha channeling to ease p-$^{11}$B ignition, including the drop in electron temperature relative to ion temperature and the higher reactivity of the fast ions. These simplifications come with three caveats.

First, we neglected the $\phi$-dependence of $\kappa$ in Eq.~(\ref{eqn:phiSolution}), and approximated the thermal proton and boron temperatures as equal.
This simplification does not significantly affect the results (see Fig.~\ref{fig:phi}), since $\phi$ remains small, and since the proton-boron thermalization is relatively fast. The full $\phi$-dependence is included in the numerical results in Fig.~\ref{fig:PLs}.
 
Second, we treated the channeling-supported fast protons as a monoenergetic population.
This approximation is not necessary in practice, since in principle the fast protons can come from the tail of the thermal proton population, diffused by  waves to higher energy near the fusion power maximum, precisely selected and energized by means of a resonance condition.

Third, we  constrained our optimization more than strictly necessary, by holding the ion temperature and the ratio of protons to boron ions fixed.
In a full optimization, that ratio would vary along with $\chi$ and $\eta$, since the optimal ion mix for one beam fraction might not be optimal for another.
Such a  fuller optimization, incorporating the unsimplified power balance equations and optimizing over all free parameters, broadly confirms the results of the simplified and compact analysis here. \cite{Ochs2022}
%

In addition to the above caveats concerning our analysis, it must be emphasized that p-$^{11}$B fusion itself comes with very significant technological challenges. Producing and confining a plasma at the required temperatures is an enormously difficult problem, and much harder than achieving DT fusion. Also, achieving large alpha-channeling efficiencies is a difficult and device-specific physics problem.\cite{Fisch1992, Fisch1995a, Ochs2015a, Ochs2015b, Herrmann1997, Cianfrani2018, Cianfrani2019, Romanelli2020, White2021} Moreover, although certain physical processes underlying the alpha channeling effect have been verified experimentally,\cite{Darrow1996, Fisch2000, Magee2019} since alpha particles have not been produced in great numbers even in DT fusion devices, the full effect itself has not yet been observed. Nonetheless, though difficult, p-$^{11}$B fusion does offer huge benefits: non-radioactive plentiful reactants, no radioactivity-producing neutrons, and no need to breed tritium.\cite{Manheimer2020} We demonstrate compactly that, including alpha channeling, at least one principal barrier to p-$^{11}$B fusion, namely producing more fusion power than is lost by bremsstrahlung and thermal conduction, might be appreciably overcome in a hybrid beam-thermal approach. This could open up new opportunities for clean, abundant fusion power.

\textbf{Acknowledgements:} 
The authors thank Mikhail Mlodik and Tal Rubin for helpful conversations. 
This work was supported by ARPA-E Grant DE-AR0001554.
This work was also supported by the DOE Fusion Energy Sciences Postdoctoral Research Program, administered by the Oak Ridge Institute for Science and Education (ORISE) and managed by Oak Ridge Associated Universities (ORAU) under DOE contract No. DE-SC0014664.

\section*{Data Availability Statement}

The data that support the findings of this study are available from the corresponding author upon reasonable request.

%

\appendix
\section{Numerical Model} \label{appendix:model}

In the main paper, Fig.~\ref{fig:PLs} is based on a numerical solution of the steady state based on Eqs.~(\ref{eq:dUfdt}), (\ref{eq:dUidt}), and (\ref{eq:dUedt}), that is: 
\begin{align}
&0 = \alpha_f P_F - (\kappa_i + \kappa_e) \phi \\
&0 = \alpha_i P_F + \kappa_i \phi + K_{ie} (T_e - T_i) -  P_L \\
&0 = \alpha_e P_F + \kappa_e \phi + K_{ie} (T_i - T_e)  - P_B. 
\end{align}
Each simulation is solved for a different choice of $\eta$ and $\chi$.
All simulations use a boron density of $1.5 \times 10^{13} \text{ cm}^{-3}$ and a total proton density of $8.5 \times 10^{13} \text{ cm}^{-3}$. 
To solve this system of equations numerically, it is necessary to include a model for the collisional slowing-down fractions $\alpha_{i0}$ and $\alpha_{e0}$; the per-particle fusion rates $y_f$ and $y_p$; the slowing-down rates $\kappa_i$ and $\kappa_e$; the temperature equilibration rate $K_{ie}$; and the bremsstrahlung power density $P_B$. 

The approach taken here has been to calculate each of these parameters for a scenario closely corresponding to the thermonuclear breakeven in Ref.~\cite{Putvinski2019}, then to allow $T_e$ and $\phi$ to vary as functions of $\eta$ and $\chi$ and to scale the coefficients as appropriate, taking the ion temperature to be fixed. 

Let $T_{e0} \doteq 160 \text{ keV}$. Then set the following: 
\begin{align}
&y_f = 1.061 \times 10^{19} \text{ eV cm}^{-3} \text{ s}^{-1} \\
&y_p = 0.534 \times 10^{19} \text{ eV cm}^{-3} \text{ s}^{-1}\\
&\tilde{\alpha}_{p0} = 0.463 \\
&\tilde{\alpha}_{b0} = 0.400 \\
&\tilde{\alpha}_{e0} = 0.138 \\
&\tilde{\kappa}_b = 1.598 \times 10^{19} \text{ eV cm}^{-3} \text{ s}^{-1} \\
&\tilde{\kappa}_p = 2.269 \times 10^{19} \text{ eV cm}^{-3} \text{ s}^{-1} \\
&\tilde{\kappa}_e = 0.529 \times 10^{19} \text{ eV cm}^{-3} \text{ s}^{-1}\\
&\tilde{K}_{ie} = 3.08 \times 10^{13} \text{ cm}^{-3} \text{ s}^{-1} .
\end{align}
These values correspond to the parameters evaluated at the thermonuclear breakeven operating regime described in Ref.~\cite{Putvinski2019}. 
For these particular parameters, $\phi = 0$, $T_b = 299 \text{ keV}$, and $T_p = 309 \text{ keV}$. 
Then, given this operating point, it is possible to evaluate quantities that are known at $T_e = T_{e0}$ and $\phi = 0$ at other values of $T_e$ and $\phi$. 
This is done as follows: 
\begin{align}
&P_F(\phi) = \phi y_f + (1-\phi) y_p \\
&\alpha_{e0} = \tilde{\alpha}_{e0} \bigg( \frac{T_e}{T_{e0}} \bigg)^{-3/2} \bigg( 1 + \frac{\tilde{\alpha}_{p0} \phi}{\tilde{\alpha}_{b0} + \tilde{\alpha}_{p0}} \bigg) \label{eqn:alphaExpansion} \\
&\alpha_{i0} = 1 - \alpha_{e0} \\
&\alpha_f = \eta \chi \\
&\alpha_i = \eta (1-\chi) + (1-\eta) \alpha_{i0} \\
&\alpha_e = (1-\eta) \alpha_{e0} \\
&\kappa_i = \tilde{\kappa}_b + (1-\phi) \tilde{\kappa}_p \\
&\kappa_e = \tilde{\kappa}_e \bigg( \frac{T_e}{T_{e0}} \bigg)^{-3/2} \\
&K_{ie} = \tilde{K}_{ie} \bigg( \frac{T_e}{T_{e0}} \bigg)^{-3/2} . 
\end{align}
Eq.~(\ref{eqn:alphaExpansion}) is an expansion in small $\phi$ and $\alpha_{e0}$. 
For any given $T_e$, $P_B$ can be evaluated using the expression found in Ref.~\cite{Putvinski2019}. 

\end{document}